\documentstyle[sprocl]{article}

\def\Journal#1#2#3#4{{#1} {\bf #2}, #3 (#4)}

\def\PLB{{\em Phys. Lett.}  B}
\def\PRL{\em Phys. Rev. Lett.}
\def\PRD{{\em Phys. Rev.} D}

\def\be{\begin{equation}}
\def\ee{\end{equation}}
\def\bea{\begin{eqnarray}}
\def\eea{\end{eqnarray}}

\begin{document}

\title{NEUTRINO MASSES AND MIXING
FROM SUPERSYMMETRIC INFLATION}

\author{ G.LAZARIDES }

\address{Physics Division , School of Technology,Aristotle University of Thessaloniki,
\\ Thessaloniki GR 540 06, Greece.}

\maketitle\abstracts{
A supersymmetric model based on a left-right symmetric 
gauge group is proposed where hybrid inflation, baryogenesis and neutrino oscillations 
are linked.This scheme, supplemented by a familiar ansatz for the 
neutrino Dirac masses and mixing of the two heaviest families and with 
the MSW resolution of the solar neutrino puzzle, implies that
$1\ {\rm {eV}\stackrel{_{<}}{_{\sim }}m_{\nu _{\tau }}
\stackrel{_{<}}{_{\sim }}9\ eV}$.
The mixing angle $\theta_{\mu \tau }$ is predicted to 
lie in a narrow range which will be partially tested by the
Chorus/Nomad experiment.}

\par
We consider a supersymmetric model based on the left-right symmetric gauge group
$G_{LR} = SU(3)_{c} \times SU(2)_{L} \times SU(2)_{R} \times U(1)_{B-L}$.
Of course, it is anticipated that $G_{LR}$ is embedded in a grand unified theory such as
$SO(10)$ or $SU(3)_c \times SU(3)_L \times SU(3)_R$. The breaking of $G_{LR}$ to the
Standard Model group is achieved by the renormalizable superpotential
$W=\kappa S (\bar{\phi}\phi - M^2),~\kappa>0, M>0$,
where $S$ is a gauge singlet left handed superfield and $\phi, \bar{\phi}$ the
Standard Model singlet components of a conjugate pair of $SU(2)_R \times
U(1)_{B-L}$ doublet left handed superfields.The superpotential $W$ leads~\cite{dss}
to hybrid inflation in a `natural' way.This means that a)there is no need of very small 
coupling constants, 
b) $W$ is the most general renormalizable superpotential allowed by the gauge and an $R$ 
symmetry, and c) supersymmetry guarantees that the radiative corrections do not
invalidate inflation. They rather provide a slope along the inflationary
trajectory which drives the inflaton towards the supersymmetric vacua. An
important parameter of the scheme is  $x_{Q} = S_{Q}/M$, where $S_{Q}$ is the
value of $S$ when our present horizon crossed outside the inflationary horizon.
We find that $1 \leq x_Q \stackrel{_<}{_\sim} 2.6 $ (see later) and, for definiteness, 
we take $x_Q\approx 2$.
From COBE one obtains $M \approx 5.5 \times 10^{15} $GeV and $\kappa \approx 4.5 
\times 10^{-3}$.

After the end of inflation, the system falls towards the supersymmetric vacua, oscillates
about them and finally decays `reheating' the universe.The `inflaton' (oscillating system)
consists of two complex scalar fields, $S$ and $\theta=(\delta\phi+\delta\bar{\phi})/
\sqrt{2}$, with $\delta\phi=\phi-M$, $\delta\bar{\phi}=\bar{\phi}-M$ and mass 
$m_{infl}=\sqrt{2}\kappa M$.$S$ can decay first through the superpotential coupling
$Sh^{(1)}h^{(2)}$, where $h^{(1)}$, $h^{(2)}$ are the higgs superfields coupled to
up and down type quarks respectively.So we concentrate on the decay of $\theta$.
The relevant coupling is given \cite{lss} by the non-renormalizable
effective superpotential term $\delta W=(M_{\nu^c}/2M^{2})\bar{\phi}\bar{\phi}\nu^{c}
\nu^{c}$, where $M_{\nu{c}}$ is the Majorana mass of the relevant right handed neutrino 
$\nu^{c}$.
The field $\theta$ decays predominantly to the heaviest $\nu^{c}$ with
$M_{\nu^{c}} \leq m_{infl}/2$.The `reheating' temperature, assuming the MSSM spectrum, 
is found \cite{lss} from the decay width of $\theta$, $\Gamma_{\theta} \approx
(1/16\pi)(\sqrt{2} M_{\nu^{c}}/M)^{2} m_{infl}$, to be $T_{R} \approx M_{\nu^{c}}/
9.23$.We will assume that $T_{R}$ is restricted by the gravitino constraint, $T_{R}
\stackrel{_<}{_\sim} 10 ^9 $ GeV.Note that
$T_{R}$ is closely linked to the mass of the heaviest $\nu^{c}$ satisfying
$M_{\nu^{c}} \leq m_{infl}/2$.

In order to obtain~\cite{neu} information about the light neutrino masses and mixing of the
two heaviest families, we ignore the first family assuming it has small mixings.
The relevant `asymptotic' (at $M_{GUT}$) $2 \times 2$ mass 
matrices are $M^{L}$, the mass matrix of charged leptons ($L^{c}$, $L$), $M^{D}$, the
Dirac mass matrix of neutrinos ($\nu^{c}$, $\nu$), and $M^{R}$, the Majorana mass matrix
of $\nu^{c}$ 's.We shall first diagonalize
$M^{L}$, $M^{D}$:
\begin{equation}
M^{L}\rightarrow M^{L}\,^{\prime }=\tilde{U}^{L^{c}}M^{L}U^{L}=\left( 
\begin{array}{cc}
m_{\mu } &  \\ 
& m_{\tau }
\end{array}
\right) \ \ ,\label{eq:lep}
\end{equation}
\begin{equation}
M^{D}\rightarrow M^{D}\,^{\prime }=\tilde{U}^{\nu ^{c}}M^{D}U^{\nu }=\left( 
\begin{array}{cc}
m_{2}^{D} &  \\ 
& m_{3}^{D}
\end{array}
\right) \ ,\label{eq:neut}
\end{equation}
where the diagonal entries are positive. This gives rise to the `Dirac'
mixing matrix $U^{\nu }\,^{\dagger }U^{L}$ in the leptonic charged currents.
Using the remaining phase freedom, we can bring this matrix to the form 
\begin{equation}
U^{\nu }\,^{\dagger }U^{L}\rightarrow \left( 
\begin{array}{cc}
{\cos \theta }^{D} & {\sin \theta }^{D} \\ 
-{\sin \theta }^{D} & {\cos \theta }^{D}
\end{array}
\right) ,\label{eq:dirac}
\end{equation}
where $\theta ^{D}(0\leq \theta ^{D}\leq \pi /2)$ is the `Dirac' (not the
physical) mixing angle in the $2$-$3$ leptonic sector.
In this basis, the Majorana mass matrix can be written as 
$M^{R}=U^{-1} M_{0} \tilde{U}^{-1}$, 
where $M_{0}=diag(M_{2}, M_{3})$, with $M_{2}, M_{3}$ (both positive) being
the two Majorana masses, and $U$ is a unitary matrix which can be
parametrized as 
\begin{equation}
U=\left( 
\begin{array}{cc}
{\cos }\theta & {\sin }\theta \ e^{-i\delta } \\ 
-{\sin }\theta \,e^{i\delta } & {\cos }\theta
\end{array}
\right) \left( 
\begin{array}{cc}
e^{i\alpha _{2}} &  \\ 
& e^{i\alpha _{3}}
\end{array}
\right) \ ,\label{eq:umatrix}
\end{equation}
with $0\leq \theta \leq \pi /2$ and $0\leq \delta <\pi $. The light neutrino
mass matrix is 
\begin{equation}
m=-\tilde{M}^{D}\,^{\prime }\ \frac{1}{M^{R}}\ M^{D}\,^{\prime }\ \
=\left( 
\begin{array}{cc}
e^{i\alpha _{2}} &  \\ 
& e^{i\alpha _{3}}
\end{array}
\right) \Psi (\theta ,\delta )\left( 
\begin{array}{cc}
e^{i\alpha _{2}} &  \\ 
& e^{i\alpha _{3}}
\end{array}
\right) \ \ ,\label{eq:mass}
\end{equation}
where $\Psi (\theta, \delta )$ depends also on $M_{2}$, $M_{3}$, $m_{2}^{D}$, 
$m_{3}^{D}$.

We will denote the two positive eigenvalues of the light neutrino mass
matrix by $m_{2}$ (or $m_{\nu _{\mu }}$), $m_{3}$ (or $m_{\nu _{\tau }}$).
Recall that all the quantities here (masses,
mixings) are `asymptotic'.The determinant
and the trace invariance of $m^{\dagger }m $ provide us with two constraints on the 
(asymptotic) parameters: 
\begin{equation}
m_{2}m_{3}\ =\ \frac{\left( m_{2}^{D}m_{3}^{D}\right) ^{2}}{M_{2}\ M_{3}}\ \
,\label{eq:det}
\end{equation}
\begin{equation}
m_{2}\,^{2}+m_{3}\,^{2}\ =\frac{\left( m_{2}^{D}\,\,^{2}{\rm c}%
^{2}+m_{3}^{D}\,^{2}{\rm s}^{2}\right) ^{2}}{M_{2}\,^{2}}+
\label{eq:trace}
\end{equation}
\[
\ \frac{\left( m_{3}^{D}\,^{2}{\rm c}^{2}+m_{2}^{D}\,^{2}{\rm s}^{2}\right)
^{2}}{M_{3}\,^{2}}+\ \frac{2(m_{3}^{D}\,^{2}-m_{2}^{D}\,^{2})^{2}{\rm c}^{2}%
{\rm s}^{2}\,{\cos 2\delta }}{M_{2}\,M_{3}}, 
\] 
where $\theta $, $\delta $ are defined in Eq.~\ref{eq:umatrix} , ${\rm c}=\cos \theta \,
,\ {\rm s}=\sin \theta $.Note that the phases $\alpha_{2},\alpha_{3}$ in 
Eq.~\ref{eq:mass} cancel out in these constraints and, thus, remain undetermined.

The mass matrix $m$ is diagonalized by a unitary rotation $V$ on $\nu$'s: 
\begin{equation}
V\ =\ \left( 
\begin{array}{cc}
e^{i\beta _{2}} &  \\ 
& e^{i\beta _{3}}
\end{array}
\right) \left( 
\begin{array}{cc}
{\cos }\varphi & {\sin }\varphi \,e^{-i\epsilon } \\ 
-{\sin }\varphi \,e^{i\epsilon } & {\cos }\varphi
\end{array}
\right) \ ,\
\label{eq:unit} 
\end{equation}
where $0\leq \varphi \leq \pi /2\ ,\ 0\leq \epsilon <\pi $.The `Dirac'
mixing matrix in Eq.~\ref{eq:dirac} is now multiplied by $V^{\dagger }$ on the left and, 
after phase absorptions, takes the form 
\begin{equation}
\left( 
\begin{array}{cc}
{\cos }\theta _{23} & {\sin }\theta _{23}\,e^{-i\delta _{23}} \\ 
-{\sin }\theta _{23}\,e^{i\delta _{23}} & {\cos }\theta _{23}
\end{array}
\right) \ ,\ 
\label{eq:mixing}
\end{equation}
where $0\leq \theta _{23}\leq \pi /2\,\ ,\ 0\leq \delta _{23}<\pi $. Here, $%
\theta _{23}$ (or $\theta _{\mu \tau }$) is the physical mixing angle in the 
$2$-$3$ leptonic sector and its cosine equals the modulus of the complex
number 
${\cos }\varphi \,{\rm {\cos }\theta }^{D}+{\rm {\sin }\varphi \sin \theta }%
^{D}\,e^{i(\xi -\epsilon )}\ ,{\rm \ }$
where $-\pi \leq \xi -\epsilon =\beta _{2}-\beta _{3}-\epsilon \leq \pi $.
The phases $\beta_{2}$,$\beta_{3}$ and 
$\xi$ remain undetermined due to the arbitrariness of $\alpha_{2}$,$\alpha_{3}$.
Thus, the precise value of $\theta _{23}$ %
cannot be found. However, we can determine the range in which $%
\theta _{23}$ lies: 
$|\,\varphi -\theta ^{D}|\leq \theta _{23}\leq \varphi +\theta ^{D},\ {\rm {%
for}\ \varphi +\theta }^{D}\leq \ \pi /2\,\cdot$

\par
We now need to know the asymptotic values of $m^{D}_{2,3}$ ,~$\theta^{D}$.
Approximate $SU(4)_{c}$ - invariance in the up quark and neutrino sectors gives 
$m^{D}_{2}=m_{c}$,~$ m^{D}_{3} = m_{t}$, sin$\theta^{D}= \mid V_{cb}\mid$ 
`asymptotically'.
Renormalization of light neutrino masses and mixing between $M_{GUT} $ and $ M_{Z}$
is also included~\cite{blp} assuming the MSSM spectrum and large 
tan$\beta \approx m_{t}/m_{b}$. In the framawork of `hierarchical' light neutrino masses 
$(m_{3} \gg m_{2} \gg m_{1})$, the small angle MSW resolusion of the
solar neutrino puzzle implies~\cite{s} $1.7 \times 10^{-3}$ eV $\stackrel{_<}{_\sim}
m_{2} \stackrel{_<}{_\sim} 3.5 \times 10^{-3}$ eV. Finally, $m_{3}$ is restricted by 
the cosmological bound $m_3 \stackrel{_<}{_\sim}$ 23 eV (for $h \approx 0.5$).

\par
We are now ready to derive~\cite{lss,neu} usefull restrictions on $M_{2,3}$.
Assume that both $M_{2,3} \leq \frac{1}{2} m_{infl}$. Then the inflaton predominantly
decays to the heaviest of the two. The determinant condition implies
that the lowest possible value of the heaviest $M_{2,3}$ is about $10^{11}$
GeV giving $T_R \stackrel{_>}{_\sim} 10^{10}$ GeV, in conflict with the gravitino 
constraint.So we must take
$1.72 \times 10^{13}$ GeV $\approx \frac{1}{2} m_{infl} \leq M_{3} \stackrel
{_<}{_\sim} 2.5 \times 10^{13}$ GeV, 
where the upper bound comes from the requirement that the coupling constant of the
non-renormalizable term responsible for the mass of the heaviest $\nu^{c}$ does not exceed
unity.(This requirement also implies the upper bound on 
$x_{Q}$.) In summary, we see that
i) $M_{3}$ is constraint in a narrow range, and  ii) the inflaton decays to the second
heaviest $\nu^{c} $ with mass $M_{2}$.

\par
Baryons can be produced, in the present scheme, only via a primordial leptogenesis
~\cite{fy} from the decay of $\nu^{c}$ 's emerging as decay
products of the inflaton. The  lepton asymmetry is then  partially
converted into the observed baryon asymmetry of the universe by `sphaleron'
effects. The lepton asymmetry is
\begin{equation}
\frac{n_{L}}{s}=\frac{9\,T_{R}}{8\pi \,m_{infl}}\,\frac{M_{2}}{M_{3}}\,\frac{%
{\rm c}^{2}{\rm s}^{2}\ \sin 2\delta \ (m_{3}^{D}\,^{2}-m_{2}^{D}\,^{2})^{2}%
}{v^{2}(m_{3}^{D}\,^{2}\ {\rm s}^{2}\ +\ m_{2}^{D}\,^{2}{\rm \ c^{2}})}\
\cdot
\label{eq:lepton}
\end{equation}
Here $v$ is the electroweak vev at $M_{GUT}$. Renormalization 
effects should also be included. Assuming the MSSM spectrum between 
1 TeV and $M_{GUT}$, the observed baryon asymmetry $n_{B}/s$ is related~\cite{iq} 
to $n_{L}/s$ by $n_{B}/s =-\frac{28}{79}(n_{L}/s)$.


\par
We will now extract~\cite{neu} restrictions on light neutrino masses and mixing.
Take a specific allowed value of $M_{3}$ (in practice, we take
its two extreme values $\frac{1}{2} m_{infl} $ or $2.5 \times 10^{13}$ GeV).
For any pair $m_{2}$,~$m_{3}$, we use the determinant condition to evaluate
$M_{2}$ and, subsequently, $T_{R}$. The gravitino constraint ($T_{R} \stackrel
{_<}{_\sim} 10^{9}$ GeV)
then gives a lower bound in the $m_{2}$,$m_{3}$ plane.This
bound together with the MSW restriction on $m_{2}$ yields a lower bound on 
$m_{3}$,namely $m_{3} \stackrel{_>}{_\sim} 0.9 $ eV (for $M_{3} \approx 2.5 \times 
10^{13}$ GeV) or 1.3 eV (for $M_{3} \approx \frac {1}{2} m_{infl}$).
The trace condition is solved with respect to  $\delta = \delta (\theta)$, 
$0 \leq \theta \leq \pi/2$,  which is then substituted in Eq.~\ref{eq:lepton} to yield
$n_{L}/s = n_{L}/s (\theta)$. Imposing the `low' deuterium bound on $n_{B}/s$ 
(0.02 $\leq \Omega_{B}h^{2} \leq$ 0.03), we find the range 
of $\theta$ where this bound is satisfied. If such a range exists, we keep 
$m_{2}$,~$m_{3}$ as satisfying the baryogenesis
constraint. This gives an upper bound in the $m_{2}$,~$ m_{3}$ plane which together with
the MSW restriction on $m_{2}$ yields an upper bound on $m_{3}$, namely $m_{3} \stackrel
{_<}{_\sim} 5.1$ eV (for $M_{3} \approx 2.5 \times 10^{13}$ GeV) or 8.8 eV (for $M_{3}
\approx m_{infl}/2$).
The allowed area in the $m_{2}$,~$m_{3}$ plane is depicted~\cite {neu} in 
Fig.~1, where the thick solid
(dashed) line corresponds to $M_{3}=m_{infl}/2$ ($M_{3} =2.5 \times 10^{13}$ GeV).
The overall allowed range for $m_{\nu_{\tau}}$ is~\cite{neu} 
1 eV $\stackrel{_<}{_\sim} m_{\nu_{\tau}} \stackrel{_<}{_\sim}$ 9 eV, 
which is interesting for the  cold plus
hot dark matter scenario for large scale structure formation in the universe.


\par
The mixing angle $\theta_{\mu \tau}$ can also be restricted ~\cite{neu}. For
every allowed $m_{2}$,~$m_{3}$ pair and every $\theta$ satisfying the baryogenesis 
constraint we
construct $V$ in Eq.~\ref{eq:unit} and, consequently, $\varphi, \epsilon$
and the allowed range of $\theta_{\mu \tau}$,~$\mid \varphi - \theta^{D} \mid \leq
\theta_{\mu \tau} \leq \varphi + \theta^{D}$.The union of all these ranges for all
allowed $\theta$'s and $m_{2}$'s for a given $m_{3}$ gives the range of 
$\theta_{\mu \tau}$
which is allowed for this value of $m_{3}$. All these ranges for all allowed $m_{3}$'s 
constitute the allowed area
on the oscillation diagram, which is depicted~\cite{neu} in Fig.~2 
(notation as in Fig.~1) in 
confrontation to past,ongoing and planned experiments.The central part 
of this allowed area will be tested by the ongoing
SBLE at CERN, NOMAD/CHORUS.
Possibly negative result from NOMAD/CHORUS will exclude a significant part of
the allowed domains in Figs.~1,2 reducing the upper
bound on $M_{\nu_{\tau}}$ to 3.7 eV. The new CERN-SBLE(TOSCA) together with the new 
CERN-MBLE (ICARUS-JURA (600t,DIS))
will cover all our predicted area on the oscillation diagram.

\par
In summary, hybrid inflation, baryogenesis and neutrino oscillations have been linked 
in the context of a supersymmetric model based on a left-right symmetric gauge group.
Our scheme leads to stringent restriction on $m_{\nu_{\tau}}$ and 
$\theta_{\mu \tau}$ to
be tested by ongoing and planned experiments. These restrictions are derived by mainly 
`physical' arguments
(gravitino and baryogenesis constraints) supplemented by a `minimal' input from fermion 
mass matrix ansaetze (only 3 input parameters ) and experiments (MSW resolution of 
the solar neutrino problem).
The choice of the gauge group is crucial since, in this case,
$\phi$ has  the quantum numbers of $\nu^{c}$ and , thus, decays to $\nu^{c}$~'s 
producing an
initial lepton asymmetry. As a consequence, the gravitino and baryogenesis constraints 
restrict the neutrino parameters.
Supersymmetry is also crucial since together with an $R$ symmetry provides a `natural' 
frame for hybrid inflation.

\section*{Acknowledgment}
This work was supported in part by the European Commission under the Human
Capital and Mobility programme, contract number CHRX-CT94-0423.

\newpage

\begin{center}
{\large {\bf {Figure Captions}}}
\end{center}

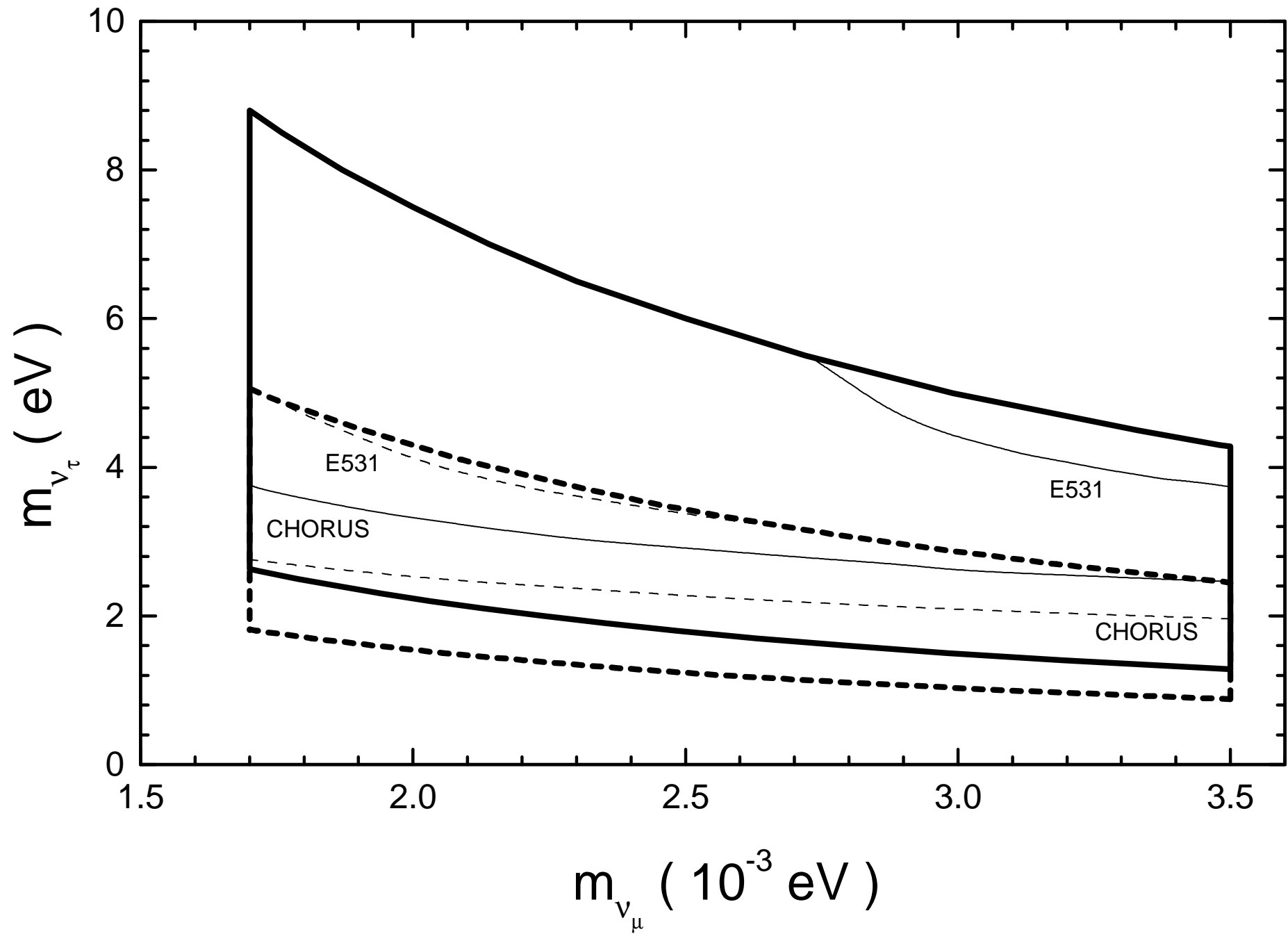
{\bf {Fig.1}.}The allowed regions in the $m_{\nu _{\mu }}$,$m_{\nu _{\tau}}$
plane.

\vspace{1cm}
{\bf {Fig.2}.}The allowed regions in the $\nu _{\mu }$-$\nu _{\tau }$
oscillation plot.


\section*{References}
\begin{thebibliography}{99}

\bibitem{dss}  G. Dvali, Q. Shafi and R.K. Schaefer, \Journal{\PRL} {73} {1886}
{1994}. For related work, see E.J. Copeland {\it  et al.}, \Journal{\PRD}{49}{6410}
{1994}; A. Linde and A. Riotto, \Journal{\PRD}{56}{1841}{1997}.

\bibitem{lss}  G. Lazarides, R.K. Schaefer and Q. Shafi,\Journal{\PRD}{56}{1324}
{1997}.

\bibitem{neu} G.Lazarides, Q. Shafi and D.N. Vlachos,{\it hep-ph}/9706385.

\bibitem{blp}  K.S. Babu, C.N. Leung and J. Pantaleone,\Journal{\PLB}{319}{191}
{1993}.

\bibitem{s}  A. Smirnov, {\it hep-ph}/9611465 and references therein.

\bibitem{fy}  M. Fukugita and T. Yanagida, \Journal{\PLB}{174}{45}{1986}; G.
Lazarides and Q. Shafi,\Journal{\PLB}{258}{305}{1991}.

\bibitem{iq}  L. Ibanez and F. Quevedo,\Journal{\PLB}{283}{261}{1992}.

\end{thebibliography}
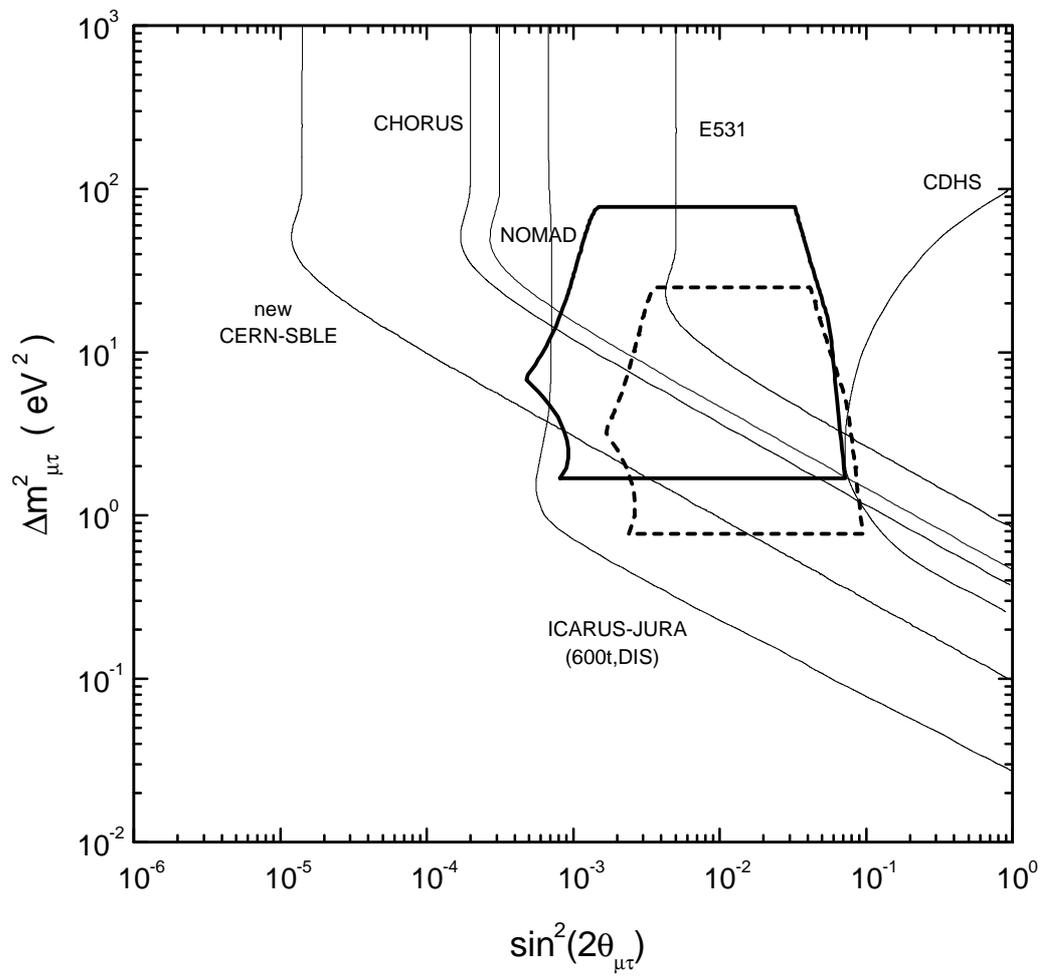
\end{document}